\documentclass[prl,aps,twocolumn,groupedaddress,showpacs]{revtex4}
\usepackage{amsmath}
\usepackage{amsbsy}
\usepackage{amssymb}
\usepackage[dvips]{graphicx}
\usepackage{epsfig}         
\usepackage{ulem}
\usepackage{color}

\newcommand{\beq}{\begin{equation}}
\newcommand{\eeq}{\end{equation}}
\newcommand{\la}{\langle}
\newcommand{\ra}{\rangle}

\newcommand{\BNm}{{\la B_N \ra}}

\renewcommand{\Sigma}{\Omega}
\newcommand{\wm}{\la D\ra}
\newcommand{\BREM}{{\cal B}_{\mbox{\tiny{REM}}}}

\begin{document}
\title{Improving free-energy estimates from unidirectional 
work measurements: theory and experiment}

\author{Matteo Palassini$^1$}\email[Corresponding author. E-mail ]{palassini@ub.edu}
\author{Felix Ritort$^{1,2}$}
\affiliation{$^1$Departament de F\'\i sica Fonamental,
Universitat de Barcelona, Diagonal 647, E--08028 Barcelona, Spain\\
$^2$CIBER-BBN de Bioingenier\'{\i}a, Biomateriales y Nanomedicina, Instituto de Salud Carlos III, Madrid, Spain}

\date{\today}

\begin{abstract}
We derive analytical
expressions for the bias of the Jarzynski free-energy estimator from
$N$ nonequilibrium work measurements, for a generic 
work distribution. To achieve this, we map the estimator onto the Random Energy Model in a suitable scaling limit parametrized by 
$(\log N)/\mu$, where $\mu$ measures the width of the lower tail of
the work distribution, and then compute the finite-$N$ corrections to
this limit with different approaches for different regimes of
$(\log N)/\mu$. We show that these expressions describe accurately
the bias for a wide class of 
work distributions, and exploit them to build an improved
free-energy estimator from unidirectional work measurements. 
We apply the method to optical tweezers unfolding/refolding experiments 
on DNA hairpins of varying loop size and dissipation,
displaying both near-Gaussian and non-Gaussian work distributions.
\end{abstract}


\maketitle 

The accurate measurement of free-energy changes has important
  applications in physics, chemistry, and biology.
Traditional measurement methods rely on reversible,
near-equilibrium transformations, which however are often unfeasible. 
In recent years, new results in nonequilibrium statistical mechanics
have suggested ways to measure free-energy changes from experiments
(and simulations) far from equilibrium (see \cite{reviewJ} for review).
The Crooks fluctuation theorem (CFT) \cite{crooks} states 
that the probability distribution $p(W)$ of the work $W$
done on a system driven out of equilibrium following an arbitrary finite-time 
protocol obeys the relation
$p(W)/p_R(-W)=e^{(W-\Delta F)/k_B T}$. Here, $p_R(W)$ is the 
work distribution (WD) for the corresponding time-reversed protocol,
$\Delta F$ is the free-energy difference between the final 
and initial equilibrium states \cite{JE}, and $T$ 
is the temperature. Hence, $\Delta F$ can 
be estimated in {\it bidirectional} experiments by 
repeating many times the forward and reverse protocol, as demonstrated 
using single-molecule manipulation techniques \cite{bustamante,collin}. 
An asymptotically unbiased estimator 
based on the CFT is the {\it acceptance ratio} (AR)  estimator \cite{shirts2004}.

In many experimental settings, 
which we shall call {\it unidirectional}, the reverse work 
cannot be measured. Examples are found in
AFM pulling of biopolymers \cite{hummer-szabo,kiang}, steered simulations
\cite{schulten}, free-energy landscape reconstruction \cite{woodside},
and single-molecule experiments on
protein unbinding, intercalation, specific cation binding, antigen-antibody interactions, 
and non-native protein conformations. 
In these cases, an alternative method
is provided by a corollary of the CFT, 
the Jarzynski equality (JE) 
$\la e^{- W/k_{\tiny {\mbox B}} T}\ra = e^{- \Delta F / k_{\tiny {\mbox B}} T}$,
where $\la\, \cdot \,\ra$ is the expectation over $p(W)$ \cite{JE}.
Given $N$
work measurements $W_1, \dots, W_N$ under the same protocol,
the Jarzynski estimator
\beq
\Delta F_N = -  \log {1\over N} 
\sum_{i=1}^N e^{- { W}_i} \,
\label{JE}
\eeq 
converges to $\Delta F$ from above as $N\to \infty$ 
(here and henceforth we set $k_{\tiny {\mbox B}} T = 1$ and
express all work values in units of $k_B T$ at room temperature).  
In practice, convergence of $\Delta F_N$ requires that rare trajectories
with $W_i < \Delta F$ be sufficiently represented,
which in turn requires $N \gg \exp({D_{typ}})$,
where $D_{typ}$ is the typical value of the 
{\it dissipated work},\/ $D = W-\Delta F$
\cite{jarzynski2006}.
Therefore, $\Delta F_N$ is a reliable estimator of $\Delta F$ 
only when $D_{typ}$ is not much larger than $k_{\tiny {\mbox B}} T$. 
It is thus important to have a quantitative estimate of the
{\it bias}\/ $B_N = \Delta F_N - \Delta F$. 
The mathematical problem faced 
is that of calculating the distribution of a (log)sum of exponentials of
i.i.d. random variables, Eq.(\ref{JE}), 
which depends on the system- and protocol-specific WD.
No closed solution to this problem
is available \cite{romeo}, even for a Gaussian 
WD (GWD). Expansions in $N^{-1}$ \cite{ZW,gore} are only applicable
when the bias is of order $N^{-1}$, i.e. smaller
than the $O(N^{-\frac{1}{2}})$ statistical error and thus negligible. 
In the relevant regime $B_N \gg O(N^{-1})$, power-law interpolations in $N$ \cite{gore}
and other approximations \cite{romeo} have 
been discussed, but no reliable analytical theory exists.

In this Letter, we derive analytical expressions for the bias expectation 
$\la B_N \ra$ for a wide class of WD's and validate them by
comparison with exact numerical simulations, also in the regime of large bias.
We use these results to build an improved unidirectional 
free-energy estimator by correcting for the bias of Eq.(\ref{JE}).
We then discuss unfolding/refolding experiments on DNA hairpins, which allow us
to test our method against the bidirectional AR estimator.

\begin{figure}[h]
\hspace{-0.5cm}
\includegraphics[height=0.9\linewidth,angle=0]{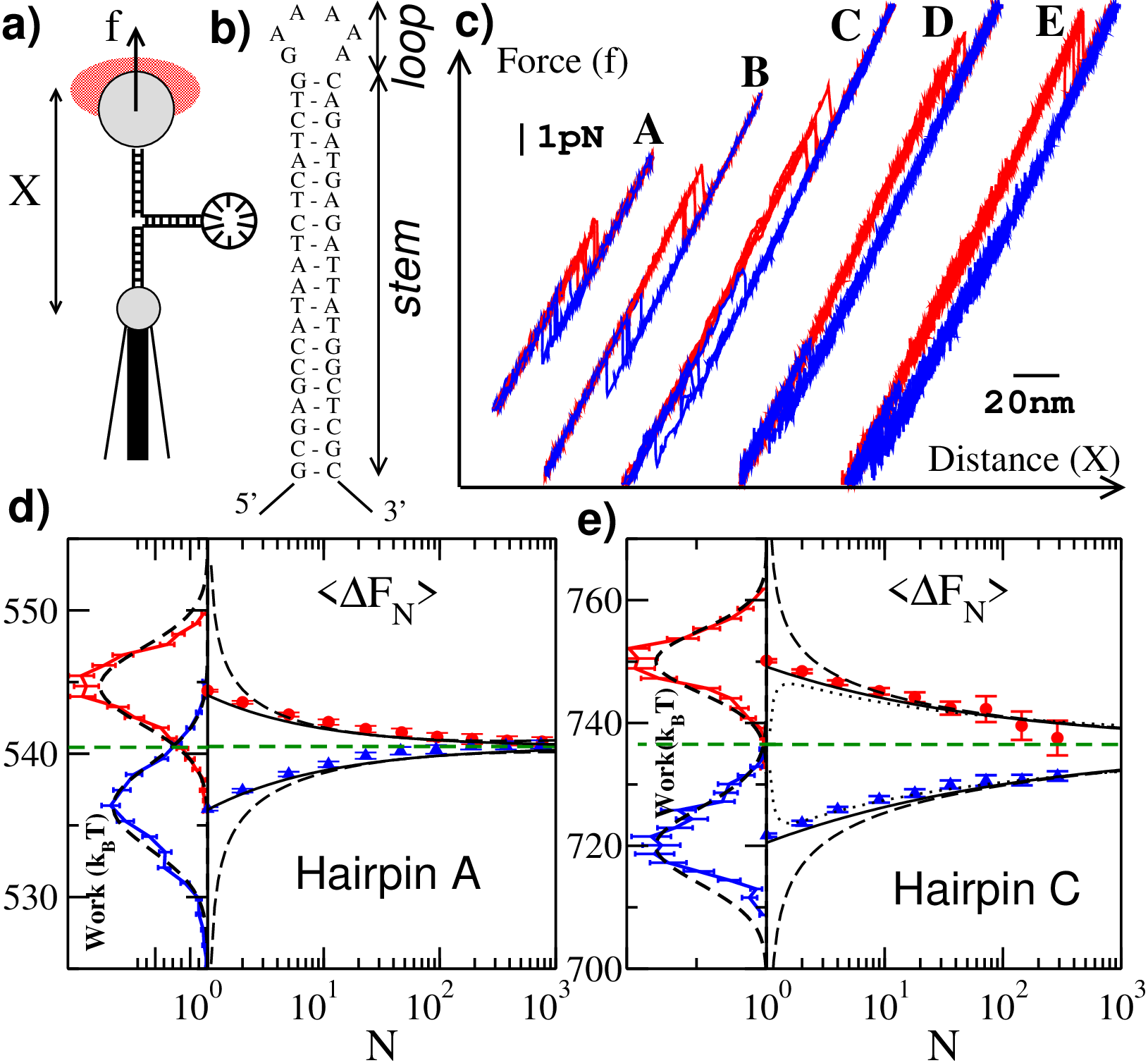}
\caption{\small {\bf Bias measurements in DNA hairpins.} 
{\bf a)} Experimental setup. 
{\bf b)} Hairpin B sequence.
{\bf c)} Examples of
  force-distance cycles for hairpins A-E. 
{\bf d)} Left:
  histograms of $p_U(W)$ (upper), $p_R(-W)$ (lower)
for hairpin A pulled at 400 nm/s. The horizontal line is the AR estimate 
$\Delta F_{AR}=540.5 \,k_B T$, giving $|\wm|=|\la W\ra -\Delta F_{AR}|$=4.3
  $k_{\tiny {\mbox B}} T$ (U), 3.9 $k_{\tiny {\mbox B}} T$
  (R). The lines are GWDs fitting the lower (upper) tail of $p_U(W)$ 
($p_R(-W)$).
Right: Jarzynski estimator $\la \Delta F_N\ra$ as a function of $N$
for U and R. Errors are estimated by jackknife. 
The lines represent $\Delta F_{AR} + \la B_N \ra$ 
with $\la B_N\ra$ given by 
  Eq.(\ref{crit}) (dashed line) and Eq.(\ref{interp}) (continuous line)  for the GWD case,
assuming $D_c=\la D \ra=\mu$, where $\mu$ is estimated from the GWD fit 
to the tails. {\bf e)} Same as d) but for hairpin C
pulled at 65 nm/s [$|\wm|$=13.6 $k_{\tiny {\mbox B}}
  T$ (U); 14.8 $k_{\tiny {\mbox B}} T$ (R)].
Also shown is $\Delta F_{AR} + \la B_N \ra$ 
with $\la B_N\ra$ given by Eq.(\ref{EV}) (dotted line).
}
\label{fig_setup}
\end{figure}

The experimental setup is shown in Fig.~\ref{fig_setup}(a).  We synthesized
five hairpins (A,B,C,D,E) with identical stem and (GAAA...) loops of
4,6,12,16,20 bases, respectively
[Fig.\ref{fig_setup}b]. The hairpins are inserted between two short (29bp)
  dsDNA handles to improve signal-to-noise resolution
  \cite{ForLorManHayHugRit11}. The construct is 
tethered to two beads, one held by
a pipette, the other by an optical trap created by
  counterpropagating laser beams \cite{Huguet10}. The light deflected by
  the trapped bead provides a direct measurement of the force acting on
  the molecule. By moving the trap away from
the pipette at constant velocity, the hairpin is stretched until it
  unfolds. Subsequent reversal of the velocity
causes the  hairpin to refold. By repeating this cycle
($\approx 200-1000$ times per experiment)
we collect the histogram of the WD's $p_{U,R}(W)$ for the work
 to unfold (U) and refold (R) the hairpin, measured by integrating the
force-distance curves (Fig.\ref{fig_setup}c) for the forward and
reverse part of each cycle (see Sec.~1 in \cite{Supp} for details). 
We divide the data in blocks of $N$ cycles, 
compute $\Delta F_N$ for each block,
 and average over the blocks to estimate $\la \Delta F_N\ra$
 for U and R separately.
As shown in  Fig.\ref{fig_setup}(d,e) for hairpins A and C, 
$\la \Delta F_N\ra$ tends to the AR estimate 
$\Delta F_{AR}$ 
for large $N$, from opposite sides for U and R.
Note that the dissipation increases with loop size and pulling speed.

We analyze theoretically the bias for a generic WD
with finite mean and an unbounded lower tail which decays as 
\beq 
p(W) \sim q {\Sigma^{\alpha-1} \over  |W-W_c|^{\alpha}}
\exp{\left(-\frac{|W-W_c|^\delta}{\Sigma^\delta}\right)}\,,
\label{pwlim}
\eeq
for $W\ll W_c$, where $W_c$ is a characteristic work value,
$\Sigma>0$ measures the tail width and $q$ is a normalization constant. 
For the JE to hold, generally one must have $\delta > 1$ \cite{tp}.
Two key parameters in the following are
\beq
\mu \equiv (\delta -1) \left({\Sigma \over\delta}\right)^{\frac{\delta}{\delta -
    1}},\,\quad \lambda \equiv
(\delta -1)\frac{\log N}{\mu}. 
\label{defs}
\eeq
A saddle point calculation
gives $\la e^{-k D} \ra \sim \exp{(\mu k^{\delta/(\delta-1)}-k
  D_c)}$ for large $\Omega$, where $D_c=W_c-\Delta F$. 
Hence the JE implies $\mu \to D_c$ in this limit. 
An example of a WD obeying Eq.(\ref{pwlim}) is a GWD with mean $\la W\ra=W_c$ and variance
$\sigma^2_W=\Sigma^2/2$ (i.e. $\alpha = 0$, $\delta=2$, $q=\pi^{-1/2}$),
for which $\la e^{-kD}\ra=\exp({\mu k^2 - k \wm})$ and thus 
$\mu=\wm=D_c=\sigma^2_W/2$ exactly for all $\Omega$. This relation
allows one to  define another unidirectional estimator 
$\Delta F_v=\la W \ra -\sigma^2_W/2$ \cite{wood}, since $\Delta F_v=\Delta F$ for the GWD.

\begin{figure}[h]
\hspace{-0.5cm}
\includegraphics[height=0.9\linewidth,angle=0]{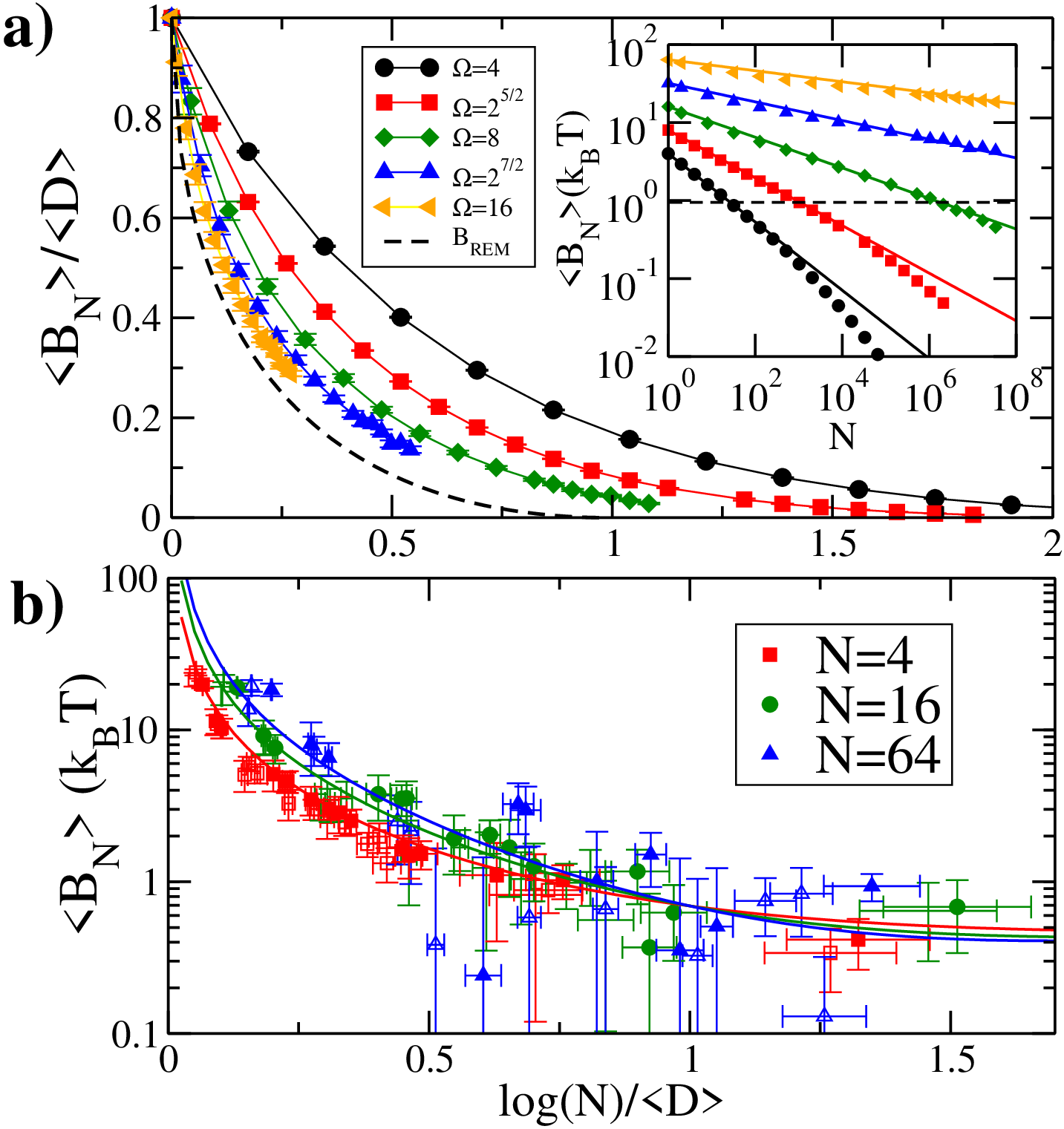}
\caption{\small  {\bf Finite-$N$ corrections to the bias for a GWD.} {\bf a)} Convergence of $\la
  B_N\ra$ to its scaling limit. The points joined by lines are averages of $B_N=-\log
  N^{-1}\sum_i e^{-D_i}$ over many sets ($D_1,\dots,D_N$) sampled
  from a GWD with variance $\Sigma^2/2$ and mean
  $\wm=\Sigma^2/4$.  The dashed line represents
  $\BREM=\mu(1-\lambda^{1/2})^2$ [(Eq.(\ref{BREM}) for $\delta=2$].
{\bf Inset of a)}: The unscaled data. The continuous lines represent
    Eq.(\ref{interp}). The horizontal dashed line
    $\la B_N \ra=1$ indicates the accuracy limit common in
    biophysical studies.  {\bf b)} Bias $|\la B_N \ra|$ 
(estimated as $|\la \Delta F_N\ra-\Delta F_{AR}|$) 
    for all experiments on hairpins A,B,C and $N=4,16,64$,
including both U and R. 
The pulling speed is in the
    range 25 - 300 nm/s and $|\wm|$ 
is in the range 1 - 20 $k_{\tiny {\mbox B}} T$.  The lines show
Eq.(\ref{crit}) for the GWD.} 
\label{fig_scaling}
\end{figure}

{\it Scaling limit --} 
Our strategy consists in computing first $\la B_N \ra$ 
in a suitable scaling limit, and then the finite-$N$
 corrections to this limit.
We obtain the scaling limit by mapping the problem onto 
the Random Energy Model (REM) \cite{derrida1} as
$B_N = D_c + \log N - \log Z_N(\beta=\Sigma/(\log_2 N)^{(\delta-1)/\delta})$, 
where  $Z_N(\beta)=\sum_{i=1}^N 
\exp[-\beta (\log_2 N)^{(\delta-1)/\delta} E_i]$
is the REM partition function and the i.i.d. variables $E_i$ 
have a distribution decaying as
$|E|^{-\alpha} \exp(-|E|^\delta)$
for $E\ll -1$ \cite{derrida1,MB}. 
From the known limit of
$N^{-1} \la \log Z_N(\beta)\ra$ for $N\to \infty$ \cite{MB,benarous}
we obtain  $\la B_N \ra \to \BREM$
in the scaling limit $(N,\Sigma)\to \infty$ with
$\lambda$ finite.
For $\lambda>1$, all 
terms in $Z_N$ give a finite contribution and we find $\BREM
=D_c - \mu$.  
For $\lambda<1$, corresponding to the glass phase of the REM \cite{MB},
$Z_N$ is dominated by a finite number of terms 
and we obtain
\beq
 \BREM = D_c + \frac{\mu(\lambda-\delta \lambda^{1/\delta})}{\delta-1} = D_c+\log N-\Sigma (\log N)^{1/\delta} \,.
\label{BREM}
\eeq
Figure~\ref{fig_scaling}a shows the approach to the scaling limit
as $\Omega$ increases, for the GWD case.
Significant deviations occur for moderate $\Omega$, 
from which the need to compute finite-$N$ corrections is apparent.

{\it Finite-$N$ corrections}  -- One must
distinguish three regimes, which require different analytical approaches: 
$\lambda > 1$, $\lambda \ll 1$, and $\lambda \lesssim 1$. 
For $\lambda > 1$, by partially resumming the $1/N$ expansion we obtain a
closed expression for $\la B_N\ra$
that improves considerably over the truncated 
expansions previously considered \cite{ZW}, which are  valid only for $\lambda \gg 1$ 
(see Sec.~2.1 in \cite{Supp}).
However, the most relevant regime in
applications of the JE is $\lambda < 1$, 
since in practice one usually has $N \ll \exp(D_{typ})$.
In the limit
$\lambda \ll 1$, using an extreme-value approach (Sec.~2.2 in \cite{Supp}) 
 we obtain to leading order 
\beq
\BNm=\BREM -\lambda^{\frac{1-\delta}{\delta}} \left[\gamma_E+
\frac{1-\alpha-\delta}{\delta} \log \log N + \log (q/\delta)\right]
\label{EV}
\eeq
where  $\gamma_E$ is the Euler-Mascheroni constant.
Cook and Derrida \cite{cook} were able to compute the finite-$N$
corrections in the critical region $\lambda \lesssim 1$ for the GWD case
in the context of the REM.
We have extended their traveling-wave approach
to the more general case of Eq.(\ref{pwlim}). 
In this way (see Sec.~2.3 in \cite{Supp} for details) we recover 
Eq.(\ref{EV}) for  $\lambda\ll 1$, while for $\lambda \lesssim 1$
we obtain
\begin{eqnarray}
\BNm &=& \BREM+\gamma_E - \lambda^{\frac{1-\delta}{\delta}} \left[\gamma_E+
\frac{1-\alpha-\frac{\delta}{2} }{\delta}\log\log N \right. \nonumber \\
&+& \left. {1\over 2}\log \frac{\pi q^2}{2\delta (\delta-1)}+\theta^2+\log 
\mbox{erfc}(\theta) \right]\,,
\label{crit}
\end{eqnarray}
where erfc is the complementary error function and
$\theta=(\lambda^{\frac{1-\delta}{\delta}}-1)\sqrt{\delta \log N}/\sqrt{2(\delta-1)}$.

{\it Numerical test --}
Equations (\ref{EV}) and (\ref{crit}) 
provide, via Eqs.(\ref{defs}) and (\ref{BREM}),  explicit expressions for
$\BNm$ as a function of $D_c, \log N$, and the shape parameters
$\delta,\Omega,\alpha,q$ of the WD tail. To illustrate the validity of these expressions,
we computed numerically $\BNm$ by sampling $W_i$ from the GWD and 
from the Weibull WD (WWD): $p(W)=\delta
\Sigma^{-\delta}|W-W_c|^{\delta-1}\exp{-(|W-W_c|/\Sigma)^\delta}$ for
$W\leq W_c$; $p(W)=0$ for $W>W_c$, where $W_c$ is fixed numerically by imposing the JE.
The WWD satisfies Eq.(\ref{pwlim}) ($\alpha=1-\delta$,
$q=\delta$) and allows us to model tails falling faster ($\delta>2$) or more slowly
($\delta<2$) than a GWD. In this case $\la W\ra = W_c - \Sigma \Gamma(1+\delta^{-1})$,
$W_{typ}=W_c-\Omega \left(\frac{\delta-1}{\delta}\right)^{1/\delta}$.

In their respective range of validity in $\lambda$, Eqs.(\ref{EV}) and (\ref{crit}) 
agree very well with the numerical data for the entire range tested ($1.1 \leq \delta \leq 3$,
$1.41 \leq \Omega \leq 16$) also for large bias,
as shown for some cases in Fig.~3(a) (for more
examples see Fig.S4 in \cite{Supp}). Furthermore, 
substituting $D_c$ with $\mu$ in Eq.(\ref{BREM}) worsens only slightly 
the agreement, an important observation for the following (see Sec.~2.5 in \cite{Supp}).
In the special case of the GWD, Eq.(\ref{crit}) gives $\la B_N\ra=\log 2$ at
$\lambda=1$. The empirical one-parameter power law 
\beq
\BNm=\wm N^{-z}, \quad z=
-\wm^{-1}\log\left(\log 2/\wm\right),
\label{interp}
\eeq
interpolating between $\lambda=1$ and $\lambda=0$ 
(for which $\la B_{N}\ra=\la D\ra$) also fits fairly well 
the GWD data, as shown in Fig.2, 
although less so than Eq.(\ref{crit}) for large $\Omega$
(see also Fig.S4 in \cite{Supp}). A power-law fit of the bias in $N$ was
proposed in Ref.\cite{gore}. 
Figure~2(b) shows that the bias of {\it all} our experiments with mild 
dissipation is well described by Eq.(\ref{crit}) for a GWD. 

\begin{figure}[h]
\hspace{-0.3cm}
\includegraphics[height=0.78\linewidth,angle=0]{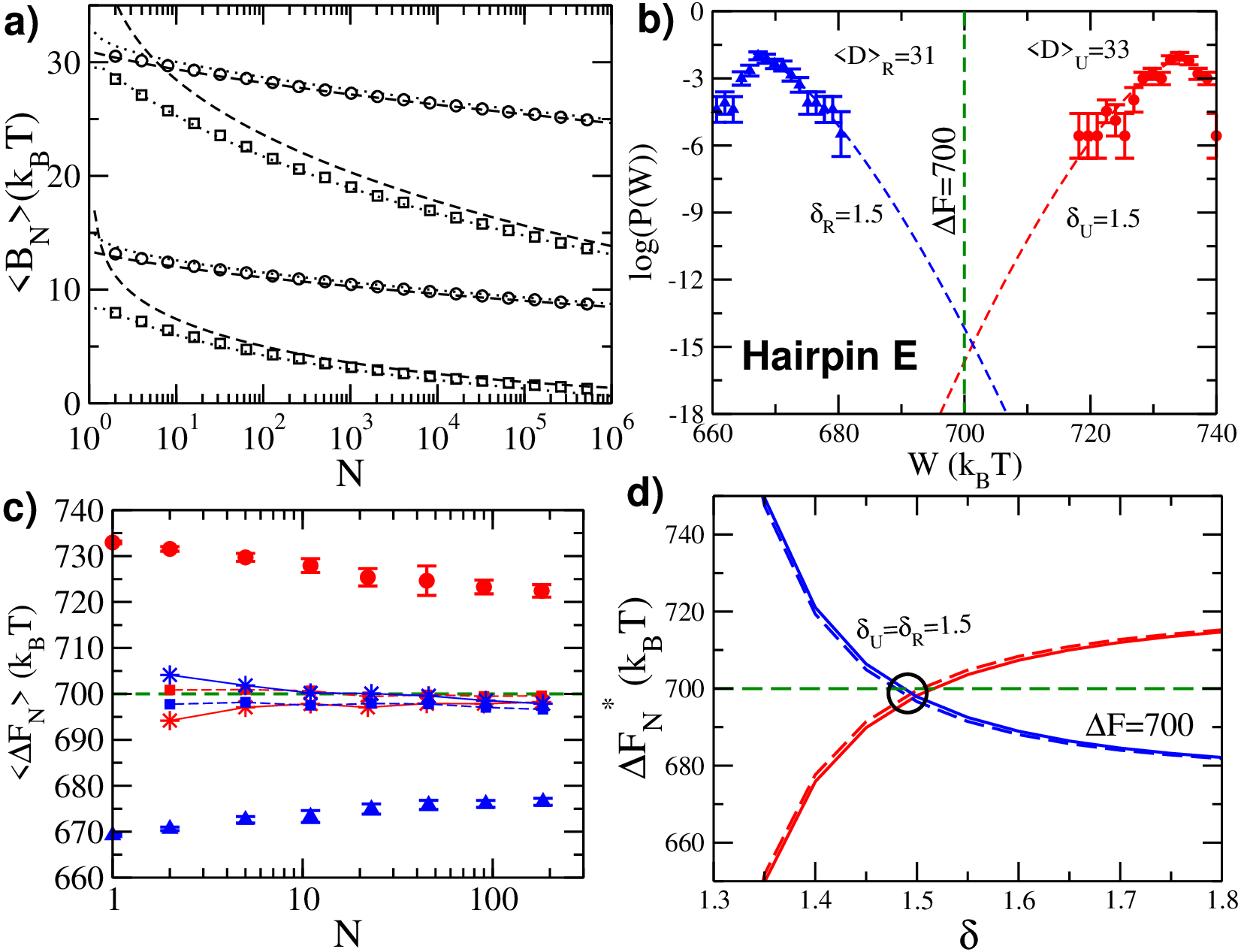}
\caption{\small  {\bf free-energy recovery for non-Gaussian WDs.}
{\bf a)} Analytical estimates of $\BNm$ as given in Eq.(\ref{EV}) (dashed lines) 
and Eq.(\ref{crit}) (dotted lines) with $D_c$ replaced by $\mu$, 
tested against simulated data for the WWD (symbols).
From top to bottom, $(\delta,\Omega,\la D \ra)$=
(1.1, 1.83, 31.0), (1.1, 1.68, 13.5), (1.5, 6, 30.1), (1.5, 4, 8.8).
{\bf b)} WD for hairpin E pulled at 130 nm/s and 
fitted WD tails assuming $\delta=1.5$ (dashed lines). 
The vertical line represents $\Delta F_{AR}$.
{\bf c)} $\la \Delta F_N\ra$ for U (circles) and R (triangles). 
Note the very slow convergence. The improved estimator $\Delta F^*_N$ obtained 
by correcting for the bias with Eq.(\ref{EV}) (squares) and Eq.(\ref{crit}) (stars)
using $\delta=1.5$ converges quickly to a value consistent
with $\Delta F_{AR}=700 k_B T$ (dashed horizontal line)
{\bf d)} $\Delta F_N^*$ recovered from U and R 
as a function of $\delta$ for $N=182$.
Using Eq.(\ref{EV}) (dashed lines) or Eq.(\ref{crit}) (continuous lines) 
gives nearly the same result.}
\label{fig_strong}
\end{figure}

{\it Free-energy recovery method --} 
The fact that Eqs.(\ref{EV}) and (\ref{crit}) describe well the bias when
$D_c = \mu$ suggests an improved estimator $\Delta F^*_N$
applicable to any problem involving the logarithm of
an exponential average:
{\it i}) given $N$ measurements $W_i$, compute $\Delta F_N$ and the
histogram of the WD; {\it ii}) estimate the
shape parameters $\Omega, \delta, \alpha, q$ (and thus $\mu$) by fitting the histogram tail to Eq.(\ref{pwlim}),
taking for instance the maximum of the WD for $W_c$ \cite{on_alpha};
{\it iii}) define $\Delta F^*_N = \Delta F_N - \la B_N \ra$ taking for $\la B_N\ra$ 
either Eq.(\ref{EV}) or (\ref{crit}) (depending on the value of $\lambda$)
and setting $D_c=\mu$.

In the special case of the Jarzynski estimator,
we must take into account the stronger constraint imposed by the CFT,
which implies that  $\la e^{-W}\ra$ is dominated by 
values of $W$ near the maximum $W^\dagger$ of the {\it reverse} WD \cite{Ritort04}. 
Sampling these very rare events is usually unfeasible, so there is no
guarantee that the fit to the measured $p(W)$ will continue to hold near $W^\dagger$.
Nevertheless, we argue that a distinction should be made 
between near-Gaussian and non-Gaussian tails. 
In the former case  (i.e. when the tail can be fitted with
$\delta \approx 2$), it is 
reasonable to assume that the fit will hold near $W^\dagger$,
hence, $\Delta F^*_N$ should give a good estimate.
The distance from a GWD can be self-consistently quantified {\it a posteriori}
by the ratio $r=\sigma^2_W/(2 \la D \ra)$ ($r=1$ for a GWD),
taking $\la D\ra = \la W \ra - \Delta F^*_N$.

We return now to our DNA experiments, for which we can compare
the unidirectional estimators $\Delta F_N, \Delta F_v, \Delta F^*_N$ 
separately for U and R with the bidirectional estimator $\Delta F_{AR}$. 
Figure 1(d) shows a case with mild dissipation, for which the bias
of $\Delta F_N$ is small and all estimators converge.
Figure 1(e) shows an experiment with intermediate dissipation.
In this case
sampling the tail of $p_U$ near the maximum of $p_R$
(or viceversa) would require an unfeasible number of cycles. 
Nevertheless, both tails
are well fitted with $\delta=2, \alpha=0$. [They are
not perfect GWD, being slightly asymmetric. 
Using $\la D\ra=\la W\ra-\Delta F_{AR}$ we obtain $r=$ 0.67 (U), 0.81 (R).]
The curves in Fig.1(e) represent 
$\Delta F_{AR} +\la B_N\ra$ with $\la B_N\ra$ given
by Eq.(\ref{EV}), (\ref{crit}), (\ref{interp}). We find that
$\Delta F_N^*$ agrees with $\Delta F_{AR}$ within
its statistical error for $N\geq 5$,
for all three expressions. 
This represents a significant 
improvement over the variance estimator, which has a bias
$\Delta F_v - \Delta F_{AR}= 4.3\pm 0.8$ (U), $-2.9 \pm  0.8$ (R),
and over the uncorrected Jarzynski estimator.
For instance,  we have $\Delta F_N-\Delta F_{AR}=
5.8\pm 3.0$ (U), $-6.6\pm 2.0$ (R) for $N=36$, and
$\Delta F_N-\Delta F_{AR}=5.1\pm 0.6$ (R) for $N=289$.
Furthermore, the fitted WD satisfy fairly well the CFT 
(see Sec.~4 in \cite{Supp}),
giving further confidence in the consistency of the method.

Finally, we consider an experiment where
the tails are far from a GWD and the dissipation is large (Fig.\ref{fig_strong}). 
In this case the WDs are wide apart
[Fig.\ref{fig_strong}(b)] and $\la \Delta F_N\ra$ converges very slowly
[(Fig.\ref{fig_strong}(c)].  
Equation (\ref{pwlim}) fits reasonably well the WD tails for a fairly
broad range of $\delta$ values (we take $\alpha=0$ for simplicity \cite{on_alpha}).
The estimator $\Delta F_N^*$ is shown in Fig.\ref{fig_strong}d:
the pronounced dependence on $\delta$ and the discrepancy between U and R 
confirm that the predictive power of
 Eqs.(\ref{EV},\ref{crit}) relies on accurately knowing $\delta$
(see Sec.~3 in \cite{Supp} for other examples).
Note also that Eqs.(\ref{EV},\ref{crit}) are ill-defined as the exponent $\delta$ approaches 
1 \cite{tp}. As $\delta$ decreases the fitted tails 
satisfy less and less the CFT (see Sec.~4 in \cite{Supp}),
signaling that $\delta$ must increase
further out in the tails.
 It is an open problem to generalize our analytical approach to an
effective $\delta$ varying with $W$.

In summary, we obtained analytical expressions for the bias of the
Jarzynski estimator, and
showed that they can be used to obtain improved unidirectional estimates
of the free energy of mechanical unfolding of DNA hairpins, 
provided the WD tail is described by
a compressed exponential over a wide enough range of work values.
These results are applicable to many unidirectional experiments
and simulations, and are relevant to other contexts 
involving sums of random exponentials.

We thank N.~Skantzos for discussions in the initial stages of this work.
MP thanks NORDITA for hospitality. Work supported by MICINN
(FIS2006-13321-C02-01, FIS2007-3454), HFSP Grant No. RGP55-2008, and ICREA Academia
grants. 


\begin{thebibliography}{99}
\bibitem{on_alpha} Since $|W-W_c|^{\alpha}$ 
is slowly varying compared to the exponential, 
in fitting Eq.(\ref{pwlim}) to experiments one can safely set $\alpha=0$.
\bibitem{tp} The JE can also be satisfied with $\delta=1,\, \Sigma<1$. This marginal
case requires a separate analysis.
\end{thebibliography}
\end{document}